\documentclass[doublecol]{epl2} 

\newcommand{\be}{\begin{equation}}
\newcommand{\ee}{\end{equation}}
\newcommand{\bea}{\begin{eqnarray}}
\newcommand{\eea}{\end{eqnarray}}

\title{Spread of wave packets in disordered hierarchical lattices}
\shorttitle{Spread of wave packets in disordered hierarchical lattices} 

\author{M. Amini\inst{1}}
\shortauthor{M. Amini}

\institute{                    
  \inst{1} Department of Physics, University of Isfahan(UI), Hezar Jerib, 81746-73441, Isfahan, Iran.
}
\pacs{05.60.Gg}{Quantum transport}
\pacs{64.60.al}{Fractal and multifractal systems}
\pacs{72.15.Rn}{Anderson or weak localization}

\abstract{
We consider the spreading of the wave packet in the generalized Rosenzweig-Porter random matrix ensemble in the region of non-ergodic extended states $1<\gamma<2$. We show that despite non-trivial fractal dimensions $0 < D_{q}=2-\gamma<1$ characterize wave function statistics in this region, the  wave packet spreading $\langle r^{2} \rangle \propto t^{\beta}$ is governed by the ``diffusion'' exponent $\beta=1$ outside the ballistic regime $t>\tau\sim 1$ and  $\langle r^{2}\rangle \propto t^{2}$ in the ballistic regime  for $t<\tau\sim 1$.  This demonstrates that the multifractality exhibits itself only in {\it local} quantities like the wave packet survival probability but not in the large-distance spreading of the wave packet.
}

\begin{document}

\maketitle

\section{Introduction}\label{Introduction}
The idea of the absence of diffusion in disordered non-interacting systems  which leads to Anderson Localization(AL) was first studied more than half a century ago~\cite{Anderson}.
Modeling the disordered  environment, Anderson considered hopping of a single particle  from atom to atom in presence of  an external random potential which can localized the particle exponentially in real space.
This matter wave localization phenomena, which is essentially a quantum mechanical effect, means that an  initial spreading of an
evolving wave-packet eventually stops in presence of sufficient amount of static disorder
and leads to an Anderson transition\cite{RMP-AT} in dimension $d>2$.

On the other hand, during the last recent years, the problem of Many-body localization(MBL)\cite{Basko} has attracted significant interests and can be understood as localization in the Fock space of Slater determinants which has hierarchical structure due to two-body interaction effect~\cite{Altshuler97}. So MBL can be studied in terms of disordered hierarchical lattices which shows an intermediate non-ergodic extended phase with multifractal eigenfunctions\cite{Luca,ALK}. Furthermore, it is also observed that  the propagation of dipolar excitations  among immobile molecules randomly spaced in a lattice  due to exchange via dipole shows the same behavior\cite{Deng}. 

Moreover, Random Matrix Theory (RMT)  which is introduced by Wigner~\cite{Wigner}  describes the statistical properties of ensembles of
random matrices.
The theory was then expanded significantly and became a major theoretical tool to calculate  the relevant statistical properties of  complex quantum systems in the field of quantum chaos~\cite{Mehta,Haake}.
However, the conventional RMT is unable to describe important phenomena such as eigenstate localization. 
For this reason, much attention has been paid to the so-called Band Random Matrices as an approach to study the AL problem theoretically\cite{F-Mirlin1991}.
Very recently~\cite{Kravtsov-NJP2015}, the existence of non-ergodic extended phase has been proven
for  a generalization of Rosenzweig-Porter(RP) RMT   suggested  in  1960~\cite{RP}.
This model also possesses the multifractality of eigenvectors in  the non-ergodic extended phase in a certain range of parameter $1<\gamma<2$.
Furthermore, it is shown  that the properties of the non-ergodic delocalized phase can be probed studying the statistics of the local
resolvent in a non-standard scaling limit~\cite{Biroli}.
So the generalized RP model seems to be the proper RMT to study the hierarchical lattices as well as statistical properties of MBL.


However, much less is clear about the spreading of wave packets 
in  hierarchical lattices   described by generalized RP model.
In this paper, we studied the spreading of  initially
$\delta$-like wave packets in the generalized orthogonal RP model of RMT, with particular attention to their behavior on different time scales.
Our calculation which is confirmed by numeric shows that  in this model  even in the non-ergodic extended phase, where the eigenstates are multifractal, there is either ballistic transport, or the diffusive one.
In the following sections, we will first introduce the model which we are interested in and
then we will describe the method which we used to monitor the wave packet spreading and finally we will report the results.

\section{Model and method}\label{Model}

One of the most interesting problems in the quantum complex systems is the problem of disordered hierarchical lattices such as the Bethe
lattice (BL) or the random regular graph (RRG).  
The complexity of these kinds of problems  is associated with the existence of an extended  non-ergodic phase which  can be addressed by a new RMT ensemble which is called generalized RP ensemble ~\cite{Kravtsov-NJP2015}.
This is an ensemble
of $N\times N$ random Hermitian matrices with  entries $H_{nm}$ 
such that each entry is an independent Gaussian random variable, 
real for orthogonal RP model and complex for the unitary RP model,
with zero mean and variance
of the off-diagonals different from that of diagonals:
\be
\langle H_{nm}\rangle=0,\hspace{.5cm} \langle |H_{nm}|^2\rangle=\left\{\begin{array}{c} 1,\quad n=m \\ \lambda^2/N^\gamma ,\quad n\neq m
      \end{array}\right.  \;,
      \label{1}
\ee
where $\lambda \sim O(N^0)$  and $\gamma$ is the main control parameter
of the model.
The orthogonal RP model (GOE), which we consider in this paper, shows three different phases.
For $\gamma>2$ all the eigenstates are completely localized and at $\gamma=2$ there is a transition from localized phase to the extended one. 
The  extended states for $1<\gamma<2$ are not ergodic with the multifractal wave functions.
In the region, the multifractal non-ergodic states are characterized by the set of non-trivial fractal dimensions $0 < D_{q}=2-\gamma<1$ for different moments of wave functions.
Finally at the $\gamma=1$ there is another transition to extended and ergodic phase.   

In comparison to RP model we may also consider the ensemble of power-law random banded matrices
 (PLBRM)\cite{Mirlin1996,Kravtsov2012,Virial1,Virial2} which is defined by
\be
\langle H_{nm}\rangle=0,\hspace{.5cm} \langle |H_{nm}|^2\rangle=\left\{\begin{array}{c} 1,\quad n=m \\ \frac{1}{2[1+(\frac{n-m}{b})^2]} ,\quad n\neq m
      \end{array}\right.  \;,
      \label{2}
\ee
where parameter $b$ controls the multifractality.

The PLBRM model as well as the RP model can be interpreted as an one-dimensional model with long-range hopping in which  the variance of hopping amplitudes decreases in a power-law manner for the former case while it does not depend on the distance but depends on the matrix size for the latter case. 
Investigating the quantum dynamics of a system which is described with such models, one may start to monitor the time evolution of a quantum state $|\psi \rangle$ which is determined
by the Schrodinger equation as

\be
      i \hbar \frac{\partial}{\partial t} |\psi(t)\rangle = H |\psi(t)\rangle,
      \label{3}
\ee
where $H$ is chosen to be of type Eqs.(\ref{1}) or (\ref{2}) for a given realization.

While the Hamiltonian operator $H$ does not explicitly depend
on time $t$, we can formally integrate the above equation in order to express the dynamics of a given quantum state.
So a full diagonalization of the Hamiltonian permits expression
of the quantum dynamics of an initial state $\psi(t_0)$ as
\be
      |\psi(t)\rangle = \sum_{n=1}^N e^{-iE_n(t-t_0)/\hbar} |n\rangle \langle n|\psi(t_0)\rangle,
      \label{4}
\ee
where $|n\rangle$ are the (time independent) eigenstates of the system and $E_n$  the corresponding eigenenergies.

To measure how a  quantum wave packet, which is located initially at site $m_0$, spreads as a function of time 
we calculate the mean square displacement, defined as
\be
   \langle r^2(t) \rangle= \langle \sum_m (m-m_0)^2 |\psi_m(t)|^2 \rangle.
      \label{5}
\ee
This quantity gives information about the extension of the wave packets and    $\langle ... \rangle$ stands for the ensemble average over different realizations of  disorder  for $H_{nm}$.

In the next section, we will use both analytical and numerical arguments to calculate the spreading of the wave packet for PLBRM and generalized RP ensembles. 

\section{Results and discussion}\label{Results}

As already mentioned above, in this section 
we  discuss the results obtained  for spreading of the wave packet in the entire multifractal phase  
of generalized orthogonal RP model.
In general, for any RMT one may considers two different regimes. In the scaling regime: 
\be
\langle r^{2}\rangle = N^{2} f(\frac{t}{N^{z}}),     \hspace{0.5cm} t>1   
      \label{6}
\ee
where the dynamic exponent $z$ is defined as the scaling correspondence
between time and volume (canonically between time and length,  $t\sim N^{z}$) with
\be
z = \left\{\begin{array}{c} 1 ,\quad $PLBRM$ \\ \gamma-1  ,\quad $RP$
      \end{array}\right.  \;.
      \label{61}
\ee
The factor $N^{2}$ in front of Eq.~(\ref{6}) is the $r^{2}$ at $t\rightarrow\infty$ and
the scaling function $f(x)$  should behave as 
\be
 f(x) \propto \left\{\begin{array}{c} $A$ ,\quad x\gg1 \\ x^\beta  ,\quad x\ll1
      \end{array}\right.  \;,
      \label{7}
\ee
where $A$ is a constant.

\begin{figure}[h!]
\onefigure[width=.98\linewidth]{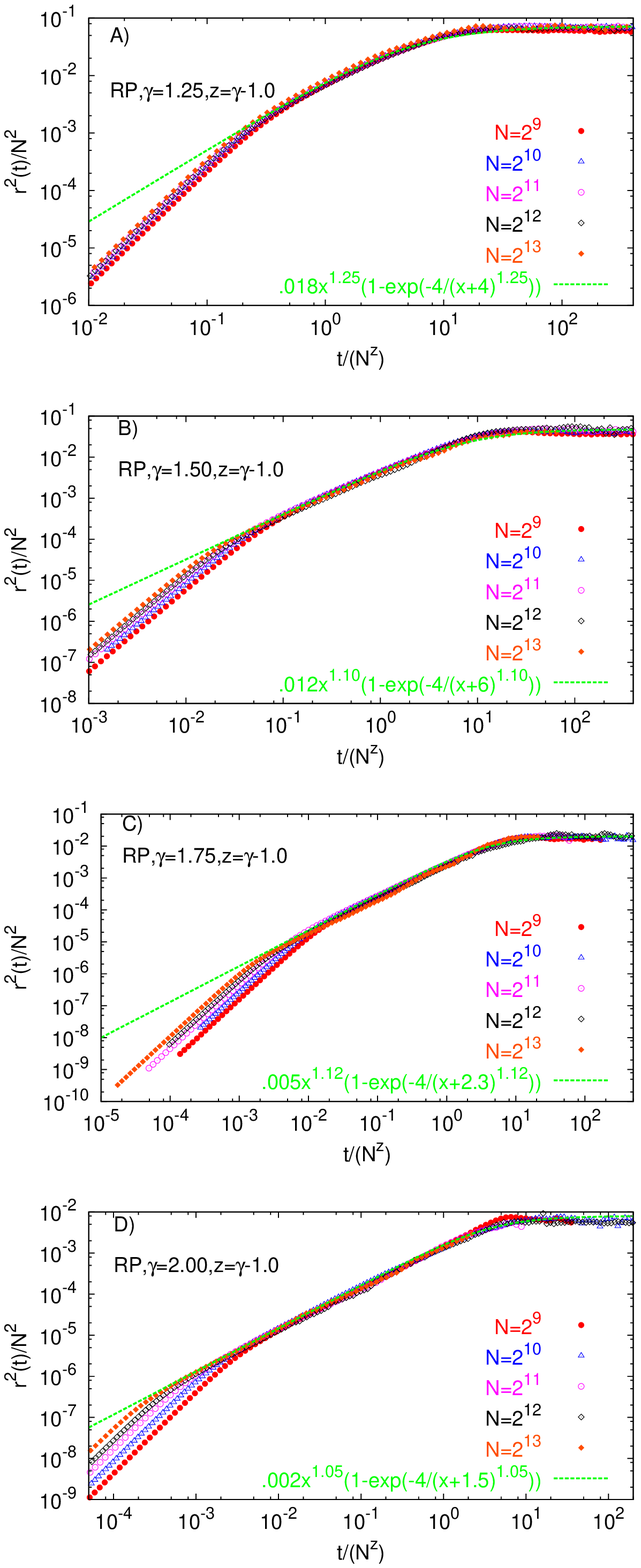}
\caption{(Color online) Data collapse of the $r^2/N^2$ vs $t/N^z$ for the orthogonal RP model with $\lambda=0.10$ and $z=\gamma-1$ at A) $\gamma=1.25$ B) $\gamma=1.50$ C) $\gamma=1.75$ and D) $\gamma=2.00$.
The green dashed line shows the best scaling function to describe the scaling
behavior.} 
\label{f1}
\end{figure}

In order to  demonstrate this scaling and find  the scaling function $f(x)$, we tried to  study numerically the time evolution of wave packets initially localized at a given site  and make  a data collapse of $r^2/N^2$ vs $t/N^z$.  
Fig.~\ref{f1} shows this scaling behavior for different ensemble sizes $N=10^9,10^{10},10^{11},10^{12},10^{13}$,
 for a system described  by generalized orthogonal RP model with $\lambda=0.10$,
 which is averaged over different realizations of disorder.
Changing the value of $\gamma=1.25,1.50,1.75,2.00$ for this model, we tried to find the best  scaling function $f(x)$ in the scaling regime which is shown
in A,B,C, and D sections respectively.

According to the asymptotic behavior of the scaling function $f(x)$ in Eq.~(\ref{7}),   we consider the following trial function: 
\be
f(x)=f_0x^\alpha [1-e^{(\frac{-S}{(x+x_0)^\alpha})}],
      \label{80}
\ee
with enough parameters to achieve a good collapse. The parameters
$f_0,$ and $x_0$ depend on $\gamma$ and the power $\alpha\approx 1.1$ and $S=4.0$ obtained for generalized RP model.

\begin{figure}[t!]
\onefigure[width=1.0\linewidth]{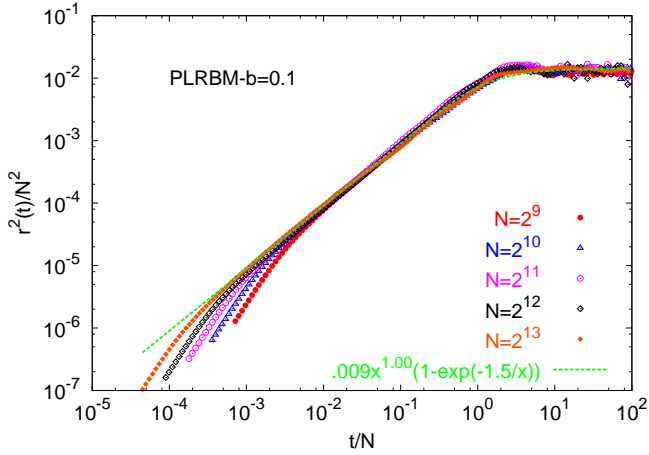}
\caption{(Color online) Data collapse of the $r^2/N^2$ vs $t/N^z$ for the PLRBM model with $b=0.1$.
The green dashed line shows the best scaling function to describe the scaling
behavior.} 
\label{f2}
\end{figure}

In comparison to RP model, we also computed the mean square displacement of the wave packet
propagation for the PLBRM ensemble which is described by Eq.~(\ref{2}). 
For this model the data collapse of $r^2/N^2$ vs $t/N^z$ is shown in 
Fig.~\ref{f2}. The best scaling function $f(x)$ in this case results to 
$S=1.5$, $x_0=0.0$ and $\alpha=1.00$.

The second regime would be the perturbative regime where $t<1$ and one my expect the ballistic transport which can be described by:
\be
\langle r^{2}\rangle = c(N) t^{2},   \hspace{0.5cm} t<1.
      \label{8}
\ee

In order to find the power $\beta$ analytically, one has to match the two regimes: the perturbative regime for $t<1$ where Eq.(~\ref{6}) does not hold and the scaling regime for $t>1$ where it is valid. According to Eq.(~\ref{6}) one obtains looking from the scaling region:
\be
\langle r^{2}(t\sim 1)\rangle \sim N^{2-\beta z} .
      \label{9}
\ee

Describing  the spreading of the wave packet at a fixed energy $\varepsilon$ in space and time, we need to calculate the generalized diffusion propagator which is the Fourier transformation of the real space density correlation function in the time domain   given by: 
\bea
D_{r,r+R}(t) &=& 2 \pi \sum_{n,m} \langle \delta(\varepsilon-\varepsilon_n)e^{\imath t(\varepsilon_m-\varepsilon_n)} \\
             && \times \psi_n(r)\psi_n^*(r+R)\psi_m(r+R)\psi_m^*(r)\rangle, \nonumber
\label{D_r}
\eea
for the regime of  $t\gg 1/B$ where $B$ is the energy bandwidth.
Here $\psi_n(r)$ and $\varepsilon_n$ are normalized wave functions and their corresponding eigenenergies respectively 
with basis set completeness which leads to $D_{r\neq r'}(t=0)=0$ and permits to replace $e^{\imath t(\varepsilon_m-\varepsilon_n)}$ by 
$e^{\imath t(\varepsilon_m-\varepsilon_n)}-1$ in the above expression.

In the regime of strong multifractality and very large distances $R=r-r'$, the wave function can be approximated as
\be
\psi_n(r)  \approx \left\{\begin{array}{c} \frac{H_{n,r}}{\varepsilon_n-\varepsilon_r} ,\quad r\neq n \\ 1  ,\quad r=n
      \end{array}\right.  \;.
\label{psi}
\ee
Also when the ratio of the off-diagonal elements $H_{nm}$ to the diagonal one $\varepsilon_n\equiv H_{nn}$ is small, which leads to concentration of wave function at one site $n$,  the eigenenergy of state $\varepsilon_n$ is almost equal to the on-site energy $H_{nn}$.   

It is now the time to consider the main contribution to Eq.~(\ref{D_r}) that comes from either $m=r$ and $n=r+R$ or $n=1$ and $m=r+R$ terms.
In both cases the combination of wave functions is equal to
\be
-\frac{|H_{r,r+R}|^2}{(\varepsilon_r-\varepsilon_{r+R})^2}
\ee 
which should be averaged over disorder. Since the averaging over diagonal entries and  over non-diagonal entries are independent, one can replace $|H_{nm}|^2$ by its average in Eqs.~(\ref{1}) and (\ref{2}) for RP and PLBRM ensemble respectively. The averaging over $\varepsilon_n$ can be done introducing the the spectral correlation function $C(\varepsilon,\varepsilon')=\langle \rho(\varepsilon)\rho(\varepsilon') \rangle$ in which $\rho(\varepsilon)$ is the averaged density of states at energy $\varepsilon$. Therefore, we have
\be
D_{r,r+R}(t)\approx 2\pi\int_{-\infty}^{\infty} d\varepsilon' C(\varepsilon,\varepsilon') \frac{1-e^{\imath t(\varepsilon-\varepsilon')}}{(\varepsilon-\varepsilon')^2} \langle |H_{nm}|^2 \rangle,
\label{Int}
\ee
with the variance of hopping matrix elements 
\be
\langle|H_{n,m}|^2\rangle = \left\{\begin{array}{c} \frac{1}{2}\frac{b^2}{R^2} ,\quad $PLBRM$ \\ \frac{\lambda^2}{N^{-\gamma}}  ,\quad $RP$
      \end{array}\right.  \;.
      \label{var}
\ee
It is worth mentioning that  the integral in Eq.~(\ref{Int}) is convergent at $\varepsilon=\varepsilon'$ due to level repulsion $C(\varepsilon,\varepsilon)=0$. At the other hand, for short time scales, when $t$ is smaller than the Heisenberg time (or $\omega\gg B/N$), the region of level repulsion is very narrow and it is possible to replace $C(\varepsilon,\varepsilon')\approx\rho(\varepsilon)\rho(\varepsilon')$. So we obtain at $\varepsilon=0$:
\be
D_{r,r+R}(t)\approx 4\pi\rho_0\langle |H_{nm}|^2 \rangle \int_{0}^{\infty} d\varepsilon' \rho(\varepsilon') \frac{1-cos(t\varepsilon')}{\varepsilon'^2},
\label{Int2}
\ee
where we used the symmetry $\rho(\varepsilon')=\rho(-\varepsilon')$ and $\rho_0$ is the density of states at zero energy.

In order to obtain the $D_{r,r+R}(t)$ for the ballistic regime in which $(\omega\sim 1/t)\gg B$, we can expand the exponent in Eq.~(\ref{Int2}) which results to
\be
D_{r,r+R}(t)\approx 2\pi\rho_0 t^2 \langle |H_{nm}|^2 \rangle \int_{0}^{\infty} dx \rho(x) .
\label{Int3}
\ee
Now, using Eq.~(\ref{var}) it is easy to see that 
\be
D_{r,r+R}(t)\sim \left\{\begin{array}{c} \frac{b^2}{R^2} t^2 ,\quad $PLBRM$ \\ N^{-\gamma} t^2  ,\quad $RP$
      \end{array}\right.  \;,
      \label{D_r_final}
\ee
which is equivalent to the results of  Refs.~\cite{Kravtsov2012,Virial1,Virial2} for the case of PLBRM.  

Finally, we calculate the mean squared diffusion length 
\be
\langle r^2(t) \rangle = \sum_{R} R^2 D_{r,r+R}(t) \sim \left\{\begin{array}{c}  b^2 N t^2 ,\quad $PLBRM$ \\ N^{3-\gamma}  t^2  ,\quad $RP$
      \end{array}\right.  \;,
      \label{r2}
\ee
and  equating this at $t=\tau\sim 1$ with Eq.~(\ref{9}) using Eq.~(\ref{61}) we find
$\beta=1$  both   for PLBRM and for RP ensembles.
Thus we prove that in the scaling region for $t<N^{z}$ there is diffusion in both cases.
The reason that the exponent $\alpha$ which is found numerically is not exactly 
the same with what we extract analytically, $\beta$, may be hidden in the finite size effects.


\section{Conclusion}\label{Conclusion}

Our conclusion is that an  initially localized wave-packet  on  disordered hierarchical lattices which is described by generalized RP RMT, spreads  either in ballistic regime, or in diffusive one. There is no non-trivial exponents $\beta$, despite that there is a non-trivial exponent $z=\gamma-1$. This is a consequence of the fact that multifractality is a local phenomenon, while $\langle r^{2} \rangle$  is determined by the second derivative of the diffusion at $k=0$, and this is non-local. This was known before and reflected in the very first paper by Chalker and Daniel on multifractality in the  quantum Hall effect at the center of Landau band~\cite{Chalker}. It is interesting that in the RMT models with multifractality it also holds.




\acknowledgments
We would like to thank V. E. Kravtsov for useful discussions and  acknowledge the $J_0$ cluster of the University of Isfahan where the computation was done.

\end{document}